\documentclass[12pt]{iopart}
\usepackage{iopams,bm,verbatim,amssymb,bm,mathptmx}
\usepackage[T1]{fontenc}
\usepackage{graphicx}
\usepackage{tikz}

\usetikzlibrary{trees}
\colorlet{lightblue}{blue!20!}
\expandafter\ifx\csname package@font\endcsname\relax\else
 \expandafter\expandafter
 \expandafter\usepackage
 \expandafter\expandafter
 \expandafter{\csname package@font\endcsname}%
\fi

\usepackage[T1]{fontenc}
\newcommand{\w}[1]{\bm{#1}} %bold math
\newcommand{\Journal}[4]{{#4} {\it #1} {\bf #2}, #3 }
\newcommand{\tho}{\textrm{\TH}}
\newcommand{\thd}{\tho '}
\newcommand{\et}{\eth}
\newcommand{\etd}{\eth '}
\newcommand{\ud}{\textrm{d}}
\newcommand{\kc}{\overline{\kappa}}
\newcommand{\lc}{\overline{\lambda}}
\newcommand{\muc}{\overline{\mu}}

\newcommand{\nuc}{\overline{\nu}}
\newcommand{\pc}{\overline{\pi}}
\newcommand{\pic}{\overline{\pi}}
\newcommand{\rhoc}{\overline{\rho}}

\newcommand{\sigmac}{\overline{\sigma}}

\newcommand{\tauc}{\overline{\tau}}

\def\H{H}
\def\g{\phi}
\def\gc{\overline{\g}}
\def\a{a}
\def\b{b}
\def\A{A}
\def\B{B}
\def\h{h}
\def\N{n}
\newcommand{\Pc}{\overline{P}} %%%%check all these!!!!!!!!!!!!!!!!!!
\newcommand{\Qc}{\overline{Q}}
\newcommand{\Nc}{\overline{\N}}
\newcommand{\xic}{\overline{\xi}}
\newcommand{\etac}{\overline{\eta}}
\newcommand{\zetac}{\overline{\zeta}}

\newcommand{\be}{\begin{equation}}
\newcommand{\ee}{\end{equation}}

\def\k{\kappa}
\def\l{\lambda}

\def\s{\sigma}

\begin{document}

\title{A new class of non-aligned Einstein-Maxwell solutions\\ with a geodesic, shearfree and non-expanding multiple Debever-Penrose vector}
\author{Norbert Van den Bergh}
\address{Ghent University, Department of Mathematical Analysis FEA16, \\ Galglaan 2, 9000 Ghent, Belgium}
\eads{\mailto{norbert.vandenbergh@ugent.be}}

\begin{abstract}
In a recent study of algebraically special Einstein-Maxwell fields\cite{NVdB2017} it was shown that, for non-zero cosmological constant, non-aligned solutions cannot have a
geodesic and shearfree multiple Debever-Penrose vector $\w{k}$. When $\Lambda=0$ such solutions do exist and can be classified, after fixing the null-tetrad such that
$\Psi_0=\Psi_1=\Phi_1=0$ and $\Phi_0=1$,  according to whether the Newman-Penrose coefficient
$\pi$ is $0$ or not. The family $\pi=0$ contains the Griffiths solutions\cite{Griffiths}, with as sub-families the Cahen-Spelkens, Cahen-Leroy 
and Szekeres metrics. It was claimed in \cite{Griffiths} and repeated in \cite{NVdB2017} that for $\pi=0$ both null-rays
$\w{k}$ and $\w{\ell}$ are non-twisting ($\bar{\rho}-\rho=\bar{\mu}-\mu=0$): while it is certainly true that $\mu (\bar{\rho}-\rho)=0$, the case $\mu=0$ 
appears to have been overlooked. 
A family of solutions is presented in which $\w{k}$ is twisting but non-expanding.
\end{abstract}


\section{Introduction}

In the quest for exact solutions of the Einstein-Maxwell equations
\begin{equation}\label{FEQS}
 R_{ab}-\frac{1}{2}R g_{ab}+\Lambda g_{ab} = F_{ac} {F_b}^c-\frac{1}{4}g_{ab}F_{cd}F^{cd},
\end{equation}
a large amount of research has been devoted to 
the study of \emph{aligned} 
Einstein-Maxwell fields, in which at least one of the principal null directions (PND's) of the electromagnetic field tensor $\w{F}$ is 
parallel to a PND of the Weyl tensor, a so called Debever-Penrose direction.
\begin{comment}One of the main properties in this respect is the Goldberg-Sachs 
theorem\cite{GoldbergSachs}, stating 
that, if a space-time admits a complex null tetrad $(\w{k},\w{\ell},\w{m},\overline{\w{m}})$ such that 
$\w{k}$ is shear-free and geodesic 
and $R_{ab}k^a k^b = R_{ab}k^a m^b = R_{ab}m^a m^b = 0$ (as is the case when $\w{k}$ is also a PND of $\w{F}$), then the Weyl tensor is algebraically special, with $\w{k}$ being a multiple
Weyl-PND. 
\end{comment}
A systematic attempt at classifying the algebraically special \emph{non-aligned} solutions was initiated in \cite{NVdB2017}. One of the topics considered, dealing with the reverse of the Goldberg-Sachs theorem,
enquired after the existence of algebraically special (non-conformally flat and non-null) Einstein-Maxwell fields with a possible non-zero cosmological 
constant for which the multiple Weyl-PND $\w{k}$ is geodesic and shear-free\footnote{Throughout I use the sign conventions and notations of \cite{Kramer} \S 7.4, with the tetrad basis vectors taken as $\w{k},\w{\ell},\w{m},
\overline{\w{m}}$ with $- k^a \ell_a = 1 = m^a
\overline{m}_a$. When using the Geroch-Held-Penrose formalism, I will write primed variables, such as $\kappa',\sigma',\rho'$ and $\tau'$, as their Newman-Penrose equivalents $-\nu,-\lambda,-\mu$ and $-\pi$.} 
($\Psi_0=\Psi_1=\kappa=\sigma=0$) 
and for which $\w{k}$ is \emph{not} parallel to a PND of $\w{F}$ ($\Phi_0\neq0$). In order to avoid frequent referring to the equations of \cite{NVdB2017} I present the 
commutator relations, GHP, Bianchi and Maxwell equations in the Appendix and I repeat part of the reasoning of \cite{NVdB2017}:\\
choosing a null-rotation about $\w{k}$ such that $\Phi_1=0$, it follows that $\Phi_2\neq 0$\footnote{with 
$\Phi_2 = 0$ $\w{\ell}$ becomes 
geodesic and shear-free and the Goldberg-Sachs theorem implies $\Psi_3=\Psi_4=0$. The Petrov type would then be D, in which case \cite{DebeverVdBLeroy,VdB1989} the only null Einstein-Maxwell solutions are given by the (doubly aligned) 
Robinson-Trautman metrics.}. Using the GHP formalism the Maxwell equations (\ref{max1},\ref{max2}) and Bianchi equations (\ref{bi1}-\ref{bi4}) become then
\begin{eqnarray}
\eth \Phi_0=0,\\
\etd \Phi_0=-\pi \Phi_0,\\
\tho \Phi_0=0,\\
\thd \Phi_0=-\mu \Phi_0,\\
\eth \Phi_2 = -\nu \Phi_0+\tau \Phi_2,\\
\tho \Phi_2=-\lambda\Phi_0+\rho \Phi_2,\\
\eth \Psi_2=-\pi \Phi_0\overline{\Phi_2}+3\tau \Psi_2,\\
\tho \Psi_2 = \mu |\Phi_0|^2+3 \rho \Psi_2,
\end{eqnarray}

after which the commutators $[\etd,\, \eth], [\etd,\, \tho], [\eth, \, \thd]$ and $[\thd,\, \tho]$ applied to $\Phi_0$ give

\begin{eqnarray}
 \eth \pi = (3\rho -\rhoc) \mu -2\Psi_2+\frac{R}{12},\label{ethpi} \\ 
 \tho \pi = 3\rho \pi, \label{dpi}\\
 \eth \mu = \lc \pi +3 \mu \tau,\\
 \tho \mu = \pi(\pc+3 \tau) +2 \Psi_2-\frac{R}{12}. \label{dmu}
\end{eqnarray}

Herewith GHP equation (\ref{ghp5d}') becomes a simple algebraic equation for $\Psi_2$,
\be
\Psi_2 = \rho \mu-\tau\pi+\frac{R}{12}, \label{psi2expr}
\ee
the $\tho$ derivative of which, using (\ref{dpi},\ref{dmu},\ref{ghp1},\ref{ghp3}), results in $\rho R=0$.\\

As $\rho =0$ would imply $\Phi_0=0$, it follows that an algebraically special Einstein-Maxwell solution possessing a shear-free and geodesic multiple Debever-Penrose vector, which is not a PND of $\w{F}$, necessarily has a vanishing cosmological 
constant. The corresponding class of solutions is non-empty, as it contains the Griffiths metrics\cite{Griffiths}, encompassing as 
special cases the metrics of \cite{CahenLeroy,CahenSpelkens,Griffiths1983,SzekeresJ1966}. 

In \cite{Griffiths} it was claimed that for $\pi=0$ both null-rays
$\w{k}$ and $\w{\ell}$ are necessarily non-twisting ($\bar{\rho}-\rho=\bar{\mu}-\mu=0$). As a consequence it was also claimed in \cite{NVdB2017} that the
Griffiths metrics are uniquely characterised by the condition $\pi=0$. However, when $\pi=0$ the only conclusion to be drawn from (\ref{ethpi}, \ref{psi2expr}) is 
that $\mu (\bar{\rho}-\rho)=0$. When $\rho$ is real this indeed leads to the metrics of \cite{Griffiths},  but the case $\mu=0$ 
appears to have been overlooked and leads, as shown in the section below, to new classes of solutions\footnote{The case $\mu=0$ is not to be regarded as a Kundt family, as the null ray generated by $\ell$ is neither geodesic nor shear-free.}.

\section{The twisting and non-expanding family}

When $\pi=0$ and $\mu=0$ the equations of the previous paragraph immediately imply $\Psi_0=\Psi_1=\Psi_2=0$ and $\Psi_3=\rho \nu -\lambda \tau$. As little progress appears to be possible in the general case, I 
restrict to solutions for which $\w{k}$ is non-expanding ($\rho+\rhoc=0$). Acting on this 
condition with the $\eth$ and $\tho$ operators, the GHP equations yield $\tau=0$ and 
\be
\rho^2+|\Phi_0|^2=0,
\ee

the $\eth$ derivative of which implies $\lambda=\Phi_2 \overline{\Phi_0} \rho^{-1}$. Translating these results into Newman-Penrose language and fixing
a boost and spatial rotation in the $\w{k},\w{\ell}$ and $\w{m},\overline{\w{m}}$ planes such that $\Phi_0=1$ and $\rho=i$, it follows that 
the only non-0 spin coefficients are $\rho$, $\nu$ and $\lambda=-i \Phi_2$, with the only non-vanishing components of the Weyl spinor being $\Psi_3=i \nu$ and $\Psi_4$.
As $[D,\, \Delta]=0$ coordinates $u,v$ and $\zeta, \zetac$ exist such that $D=\partial_u$, $\Delta=\partial_v$ and 
\be 
\delta = e^{-i u}(\xi \partial_{\zeta}+\eta \partial_{\zetac}+P \partial_u+Q\partial_v), \label{d1_def}
\ee
$\xi,\eta,P,Q$ being arbitrary functions. The $ e^{-i u}$ factor is included for convenience: applying the $[\delta,\, D]$ commutator to $u,v$ and $\zeta$ shows that 
$\xi,\eta,P,Q$ are functions of $v,\zeta,\zetac$ only. Introducing new variables $\N=e^{-iu}\nu$ and $\g=e^{-2iu} \Phi_2$ it follows that also $\N$ and $\g$ depend on $v,\zeta,\zetac$ only, with
the full set of Jacobi and field equations reducing to the following system of pde's:
\begin{eqnarray}
P_v+i \Pc\, \gc-\Nc = 0, \label{cond0}\\
Q_v+i\Qc\,\gc = 0, \label{cond1}\\
 \xi_v+i \etac \, \gc = 0, \label{cond2}\\
\eta_v+i \xic \, \gc =  0, \label{cond3} 
\end{eqnarray}
\begin{eqnarray}
 e^{-iu} \overline{\delta} P-e^{iu} \delta \Pc-2 i |P|^2 = 0, \label{cond4} \\
 e^{-iu} \overline{\delta} Q-e^{iu} \delta \Qc-2 i \Re (Q\Pc-1) = 0, \label{cond5} \\
 e^{-iu} \overline{\delta} \xi-e^{iu} \delta \etac-2 i \Re (\xi\Pc) = 0, \label{cond6}
\end{eqnarray}
\begin{eqnarray}
 e^{iu}\delta \N =-i P\N+2|\g|^2, \label{cond7} \\
 e^{iu} \delta \g = -2 i P \g-\N, \label{cond8}
\end{eqnarray}
with the $\Psi_4$ component of the Weyl spinor given by $\Psi_4=i e^{2iu}(\Nc\Pc+\Delta \g)+e^{iu}\overline{\delta}\N$. 

In the next section the case $\g=\g(\zeta,\zetac)$ will be solved explicitly. 

\section{The case $\g=\g(\zeta,\zetac)$}
%%% Follow the Maple worksheet!!!!

When $\g=\g(\zeta,\zetac)$, writing $\g=\H^2 \h^2$ with $\H>0$ and $|\h|=1$, (\ref{cond1},\ref{cond2},\ref{cond3}) integrate to $Q=q_1 e^{\H^2 v}+q_2 e^{-\H^2 v}$, 
$\xi= \overline{c_1} e^{\H^2 v}+ \overline{c_2} e^{-\H^2 v}$, $\eta= i \h^{-2} (-c_1 e^{\H^2 v}+ c_2 e^{-\H^2 v}) $ with $q_A,c_A$ depending on $\zeta,\zetac$ only and $q_J +i h^{-2} \overline{q_J}=0$ ($J=1,2$).
A coordinate transformation $\zeta \to \tilde{\zeta}(\zeta,\zetac)$ allows one then to put (writing $\tilde{\zeta}=x+i y$ and re-defining $q_J$), 
\be
\fl \delta= e^{-i u}[P \partial_u + e^{\H^2 v-C_1-i \frac{\pi}{4}}\h^{-1}(\partial_x+ q_1 \partial_v)+ e^{-\H^2 v-C_2+i \frac{\pi}{4}}\h^{-1}(\partial_y+ q_2 \partial_v)] ,
\ee
with $C_J$ and $q_J$ real functions of $x$ and $y$. 

An expression for $P$ is obtained from  (\ref{cond6}),
\begin{eqnarray}
 P= e^{i\frac{\pi}{4}} \h^{-1} [& e^{\H^2 v -C_1}(\frac{\h_{,x}}{\h}-{C_2}_{,x}-v(\H^2)_{,x}-q_1 \H^2) \nonumber\\
& +i e^{-\H^2 v -C_2}(\frac{\h_{,y}}{\h}-{C_1}_{,y}+v(\H^2)_{,y}+q_2 \H^2)],
\end{eqnarray}

after which $\N$ follows from (\ref{cond0}),
\begin{eqnarray}\label{N_def}
\N= - 2 \h\H^2 \{& e^{\H^2 v-C_1-i \frac{\pi}{4} } [ (1+2 v \H^2) \frac{\H_{,x}}{\H}+{C_2}_{,x}+q_1 \H^2] \nonumber \\
& +e^{-\H^2 v-C_2+i \frac{\pi}{4}} [ (1-2 v \H^2) \frac{\H_{,y}}{\H}+{C_1}_{,y}-q_2 \H^2] \}.
\end{eqnarray}

Herewith (\ref{cond4}) becomes a polynomial identity in powers of $v$ and $e^{\H^2 v}$, 
\be
 v^2 e^{2\H^2 v} \H_{,x}^2- v^2 e^{-2\H^2 v} \H_{,y}^2+\ldots = 0, \label{vexpv}
\ee
showing that $\H$ is necessarily constant. 

Introducing new variables 
$r_1=-\H^2q_1-{C_2}_{,x},\ r_2=\H^2 q_2-{C_1}_{,y} $
the remaining coefficients of (\ref{vexpv}) lead to the equations,
\begin{eqnarray}
 {r_1}_{,x} -2 r_1^2 -r_1 (C_1+C_2)_{,x}=0,\label{r1_eq}\\
 {r_2}_{,y} -2 r_2^2 -r_2 (C_1+C_2)_{,y}=0.\label{r2_eq}
\end{eqnarray}
Substituting this in (\ref{N_def}), equation (\ref{cond7}) becomes a Liouville equation determining $C_1+C_2$,
\be
(C_1+C_2)_{,xy}+2 \H^2 e^{C_1+C_2}=0,\label{Liouville}
\ee
while (\ref{cond8}) reduces to an identity. A final equation is (\ref{cond5}), which now becomes
\be
{r_2}_{,x}+{r_1}_{,y}-\H^2 e^{C_1+C_2}=0.\label{restriction}
\ee
The general solution of the Liouville equation being given by
\be
e^{C_1+C_2}= -\frac{\a_{,x}\b_{,y}}{\H^2(\a+\b)^2},
\ee
($\a=\a(x)$ and $\b=\b(y)$ arbitrary functions), $r_1$ and $r_2$ are given by (\ref{r1_eq},\ref{r2_eq}) as,
\begin{eqnarray}
 r_1=\frac{\b_{,y}}{2(\a+\b)(1+\A(\a+\b))},
 r_2=\frac{\a_{,x}}{2(\a+\b)(1+\B(\a+\b))},
\end{eqnarray}
with arbitrary functions $\A=\A(x),\B=\B(y)$. Herewith (\ref{restriction}) reduces to the condition
\be
(\frac{\ud \A}{\ud \a}-\A^2)(1+\B(\a+\b))^2+(\frac{\ud \B}{\ud \b}-\B^2)(1+\A(\a+\b))^2=0,
\ee
implying either $\A_{,\a}-\A^2=\B_{,\b}-\B^2=0$, or $\log \frac{1+\B(\a+\b)}{1+\A(\a+\b)}$ being separable in $x$ and $y$. As the latter condition again can be shown to imply
$\A_{,\a}-\A^2=\B_{,\b}-\B^2=0$, we conclude that the general solution is given by
\begin{eqnarray}
 r_1= \frac{\a_{,x}}{2(\a+\b)} \frac{k-k_0\b}{k+k_0 \a}\\
 r_2= \frac{\b_{,y}}{2(\a+\b)} \frac{k-k_0\a}{k+k_0 \b},
\end{eqnarray}
with either $k_0=1$ and $k$ an arbitrary (real) constant\footnote{which can be taken to be 0 or 1}, or $k_0=0,k=1$ (corresponding to the special case $\A=\B=0$). 

The resulting metric appears to contain two arbitrary functions, being the phase factor $h(x,y)$ of $\phi$ and the function 
$F(x,y)$ defined by $e^{C_1-C_2}=-\frac{\a_{,x}}{{\b}_{,y}} e^{2 F}$. These however can be eliminated by the coordinate transformations $u \to u - i \log \h$ and $v \to \H^2 v-F$, after which the dual basis takes the form,

\begin{eqnarray}
 \w{\omega}^1 &=& \frac{e^{i u}}{2\H(\a+\b)}(e^{i\frac{\pi}{4}-v}\ud \a -e^{-i\frac{\pi}{4}+v}\ud \b),\label{sol1}\\
 \w{\omega}^3 &=& \frac{1}{\H^2}\ud v -\frac{1}{\H^2(\a+\b)}[\frac{(2\a+\b)k_0+k}{2(k_0\a+k)}\ud \a -\frac{(\a+2\b)k_0+k}{2(k_0\b+k)}\ud \b],\label{sol2}\\
 \w{\omega}^4 &=& \ud u +\frac{1}{2(\a+\b)}[\frac{k_0\a-k}{k_0 \b+k}e^{-2 v} \ud \a -\frac{k_0\b-k}{k_0 \a+k}e^{2 v} \ud \b] .\label{sol3}
\end{eqnarray}

\section{Discussion}
The null tetrad (\ref{sol1}-\ref{sol3}) determines a (presumably) new family of Einstein-Maxwell solutions with zero cosmological constant and with Maxwell field and energy-momentum tensor given by 
\begin{eqnarray}
 \w{F} &=& i \H^2(\w{\omega}^1-\w{\omega}^2)\wedge \w{\omega}^3+i (e^{-2i u} \w{\omega}^1-e^{2 i u} \w{\omega}^2)\wedge \w{\omega}^4,\label{solmax}\\
 \w{T} &=& 2 \H^2 (e^{-2 i u} \w{\omega}^1 \otimes \w{\omega}^1 +e^{2 i u} \w{\omega}^2 \otimes \w{\omega}^2+ \H^2 \w{\omega}^3 \otimes \w{\omega}^3)+2 \w{\omega}^4 \otimes \w{\omega}^4.\label{solenergy}
\end{eqnarray}

The Petrov type is III, with the multiple Debever-Penrose vector $\w{k}= \partial_u$ being geodesic, shear-free and twisting but non-expanding. The real null vector $\w{\ell}$, fixed by a null-rotation such that $\Phi_1=0$, is non-diverging, 
but is non-geodesic and has non-vanishing shear. It follows that figure (2) in \cite{NVdB2017} has to be amended as in Fig.~1 below.

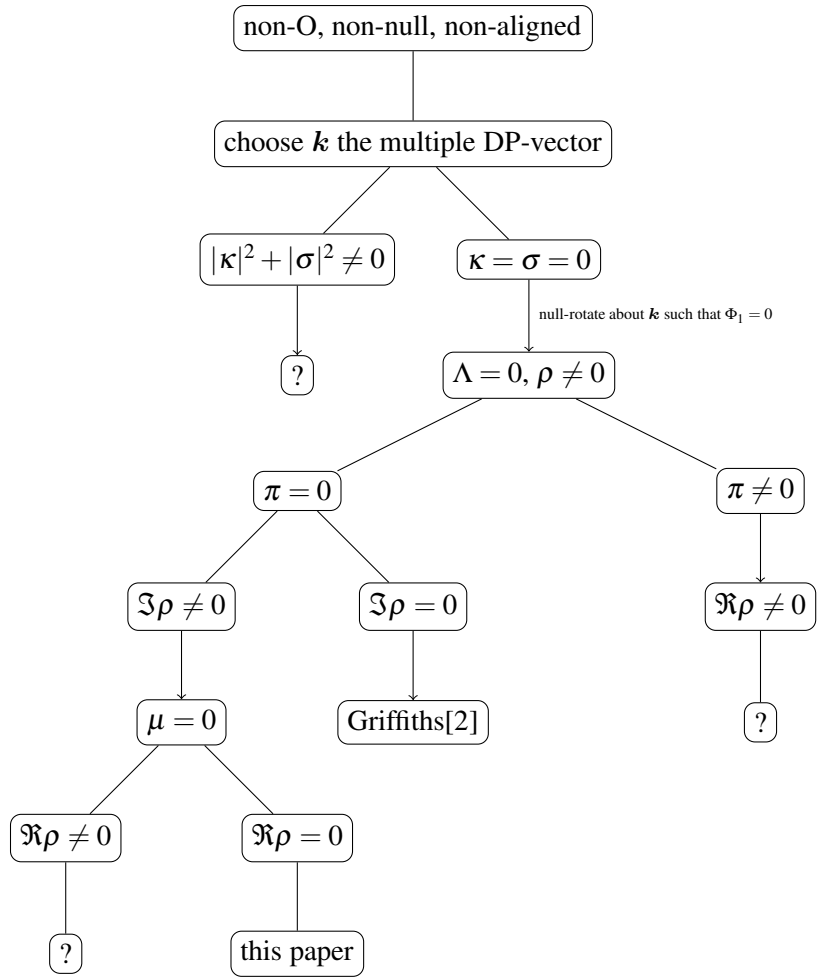
\begin{figure}[!htb]
\small{
\begin{center}
\begin{tikzpicture}[level distance=4em,
  every node/.style = {shape=rectangle, rounded corners,
    draw, align=center}]
    [\tikzstyle{level 1}=[sibling distance=15em],\tikzstyle{level 2}=[sibling distance=8em],\tikzstyle{level 3}=[sibling distance=8em]]]
    \node (a) {non-O, non-null, non-aligned}
    child { node (a0) {choose $\w{k}$ the multiple DP-vector}  
      child { node (b) {$|\kappa|^2+|\sigma|^2\neq 0$}
        child { node (b1) {?}
            edge from parent [->] node [right,draw=none] {}
        }
      }
      child { node (c) {$\kappa=\sigma=0$}
        child { node (c1) {$\Lambda = 0$, $\rho\neq 0$}
          child { node (c11) {$\pi=0$}
           child { node (c11a) {$\Im \rho \neq 0$}
             child { node (c11aa) {$\mu=0$}
               child { node (c11aa1) {$\Re \rho \neq 0$}
                 child { node (c11aa1b) {?}
                 }
               }
               child { node (c11aa2) {$\Re \rho = 0$}
                 child { node (c11aa2b) {this paper}
                 }
               }
               edge from parent [->] node [right,draw=none] {}
             }
           }
           child { node (c11b) {$\Im \rho = 0$}
            child { node (c112) {Griffiths\cite{Griffiths}}
            edge from parent [->] node [right,draw=none] {}
            }
           }
          }
          child [missing] %increases the distance between the pi=0 and pi\neq 0 children
          %child { node[right = 2cm] (c12) {$\pi\neq0$} %alternatively use this with the positioning library
          child { node (c12) {$\pi\neq0$}
            child { node (c121) [label={[draw=none]}] {$\Re \rho \neq 0$}
              child { node (c1211) {?}
              }
            edge from parent [->] node [left,draw=none] { }  
            }  
          }  
        %edge from parent [->] node [left,draw=none] {\tiny \resfour }
        edge from parent [->] node [right,draw=none] {\tiny null-rotate about $\w{k}$ such that $\Phi_1=0$}
        }
      }
    }
    ;
    \end{tikzpicture}
\end{center} 
}
\caption{\label{fig:nonaligned} {Algebraically special non-nul Einstein-Maxwell solutions for which the multiple Weyl-PND $\w{k}$ is not a PND 
of $\w{F}$.}}
\end{figure}
For all solutions $\partial_u$ is clearly a null Killing vector. While in general (i.e.~with $k$ and $k_0 \neq 0$) the isometry group
is 2-dimensional, with the second Killing vector given by
\be
\w{K}_2= k_0^2(\a+\b)\partial_v+(\a^2k_0^2-k^2) \partial_{\a} -(\b^2k_0^2-k^2)\partial_{\b}, 
\ee
the special cases $k_0=0,k=1$ and $k_0=1,k=0$ admit a 3-dimensional group of isometries, with third Killing vector
\be
\w{K}_3= \a\partial_{\a}+\b\partial_{\b}.
\ee
In the latter cases the isometry group has Bianchi type III, with the orbits being time-like hypersurfaces parametrized by the null coordinate $v$. For the case $k_0=0,k=1$ the tetrad simplifies to
\begin{eqnarray}
 \w{\omega}^1 &=& \frac{e^{iu}}{2\H(\a+\b)}[e^{i\frac{\pi}{4}-v}\ud \a-e^{-i\frac{\pi}{4}+v}\ud \b],\\
 \w{\omega}^3 &=& \frac{1}{\H^2} [\ud v -\frac{1}{2 (\a+\b)} \ud (\a-\b)],\\
 \w{\omega}^4 &=& \ud u -\frac{1}{2(\a+\b)}[e^{-2v}\ud \a-e^{2v} \ud \b]
\end{eqnarray}
and the line-element becomes
\be
\fl \ud s^2 = \frac{1}{\H^2}[-2\ud v +\frac{1}{\a+\b}\ud (\a-\b)]\ud u +\frac{1}{\H^2(\a+\b)}(e^{-2 v}\ud \a-e^{2 v}\ud \b)\ud v+\frac{\cosh{2v}}{\H^2(\a+\b)^2}\ud \a \ud \b . \label{newmetric}
\ee
The non-0 components of the Weyl-spinor are then given by
\begin{eqnarray}
 \Psi_3 = -\H^3 e^{i (u-\frac{\pi}{4})}(e^v+i e^{-v}) \\
 \Psi_4 = 2 \H^4 e^{2 i u} \cosh 2 v .
\end{eqnarray}

All the Carminati-McLenaghan invariants are regular functions of the essential coordinate $v$ over the interval 
$]-\infty,\, +\infty[$. The same holds for the case $k_0=1,k=0$ (in which the essential coordinate is $v+\log (\b / \a)$), but although all the (CM-) invariants are transformed into each other under the coordinate transformation 
$v\to v+\log (\b / \a)$ the two special cases are inequivalent. 

$\partial_u$ being Killing vector, it might look peculiar that the Weyl spinor components $\Psi_3$ and $\Psi_4$ still depend on $u$, even though
the frame was ``invariantly'' fixed. This is due to the fixation having been done by means of a null rotation putting $\Phi_0=1$: the resulting frame scalars are then not genuine Cartan invariants and, as 
the Maxwell field itself does not
inherit the space-time symmetries ($\w{F}$ is not Lie-propagated along the integral curves of the null Killing vector $\partial_u$), the frame scalars depend on $u$ as well. Note that
this also shows that the present solutions are distinct from
the Luc\'acs et al.~solutions admitting null-Killing vectors \cite{Lucacs}, as there the Maxwell field does inherit all the space-time symmetries.\\

\section*{Appendix A}
Weights\footnote{Objects $x$ transforming under boosts and rotations as
$x \rightarrow
A^{\frac{p+q}{2}}e^{i\frac{p-q}{2}\theta} x$
are called {\em well-weighted of type} $\left(p,q\right)$.} of the spin-coefficients, the Maxwell and Weyl spinor components and the GHP operators:
\begin{eqnarray} 
&\kappa : [3, 1], \nu : [-3, -1], \sigma : [3, -1], \lambda : [-3, 1], \nonumber \\
&\rho : [1, 1], \mu : [-1, -1], \tau : [1, -1], \pi : [-1, 1], \nonumber \\
&\Phi_0 : [2, 0], \Phi_1 : [0, 0], \Phi_2 : [-2, 0], \nonumber \\
&\Psi_0 : [4, 0], \Psi_1 : [2, 0], \Psi_2 : [0, 0], \Psi_3: [-2,0], \Psi_4 : [-4, 0] ,\nonumber \\
&\eth : [1, -1], \etd : [-1,1], \thd : [-1,-1], \tho : [1,1]. \nonumber
\end{eqnarray}
The prime operation is an involution with 
\begin{eqnarray}
 \kappa'=-\nu,\sigma'=-\lambda,\rho'=-\mu, \tau'=-\pi,\\
 {\Psi_0}'=\Psi_4, {\Psi_1}'=\Psi_3, {\Psi_2}'=\Psi_2,\\
 \Phi_0'=-\Phi_2, \Phi_1'=-\Phi_1.
\end{eqnarray}
%\noindent The GHP commutator relations, GHP, Maxwell and Bianchi equations ($R=4 \Lambda = constant$ and $\Phi_{IJ}=\Phi_i \overline{\Phi_j}$):\\

\noindent The GHP commutators acting on $(p,q)$ weighted quantities are given by:
\begin{eqnarray}
&[\tho,\tho'] =(\pi+\tauc)\et +(\pic+\tau)\etd +(\k\nu-\pi\tau+  \frac{R}{24}-\Phi_{11}-\Psi_2)p\,  \nonumber\\
&\quad+(\kc\nuc-\pic\tauc+  \frac{R}{24}-\Phi_{11}-\Pc_2)q, \\
&[\et,\etd] =(\mu-\muc)\tho +(\rho-\rhoc)\tho' +(\l\s-\mu\rho-  \frac{R}{24}-\Phi_{11}+\Psi_2)p\,  \nonumber\\
&\quad-(\overline{\l\s}-\muc\rhoc-  \frac{R}{24}-\Phi_{11}+\Pc_2)q,  \\
&[\tho,\et] =\pic\,\tho -\k\tho' +\rhoc\,\et +\sigma\etd+(\k\mu-\s\pi-\Psi_1)p\,  \nonumber\\
&\quad+(\overline{\k\l}-\pic\rhoc-\Phi_{01})q .  
\end{eqnarray}

\noindent GHP equations:
%NB: ghp4=GHP8, ghp5=-GHP4', ghp6=-GHP5' of our Maple package
\begin{eqnarray}
\tho\rho-\etd\k=\rho^2+\s\sigmac-\kc\tau+\k\pi+\Phi_{00}, \label{ghp1}\\
\tho\s-\et\k=(\rho+\rhoc)\s+(\pic-\tau)\k+\Psi_0, \label{ghp2}\\
\tho\tau-\tho'\k=(\tau+\pic)\rho+(\tauc+\pi)\s+\Phi_{01}+\Psi_1, \label{ghp3}\\
\tho \nu-\tho' \pi = (\pi+\tauc)\mu+(\pic+\tau)\lambda+\Psi_3+\overline{\Phi_1}\Phi_2,\label{ghp6}\\
\et\rho-\etd\s=(\rho-\rhoc)\tau+(\mu-\muc)\k+\Phi_{01}-\Psi_1,\label{ghp8}\\
\tho'\s-\et\tau=-\s\mu-\lc\rho-\tau^2+\k\nuc-\Phi_{02},\label{ghp4d}\\
\tho'\rho-\etd\tau=-\muc\rho-\l\s-\tau\tauc+\k\nu-\frac{R}{12}-\Psi_2. \label{ghp5d}
\end{eqnarray}

\noindent Maxwell equations:
\begin{eqnarray}
\tho \Phi_1-\etd \Phi_0 =\pi \Phi_0+2\rho\Phi_1-\kappa \Phi_2, \label{max1}\\
\tho \Phi_2-\etd \Phi_1 =-\lambda \Phi_0+2 \pi \Phi_1+\rho\Phi_2. \label{max2}
\end{eqnarray}

\noindent Bianchi equations (with $\Phi_{IJ}=\Phi_I\overline{\Phi_J}$ and $\Lambda=R/4=constant$):
\begin{eqnarray}
\fl {\etd} \Psi_{{0}}  -{\tho} \Psi_{{1}}
 +{\tho} \Phi_{{01}}  -{\et} \Phi_{
{00}}  =-\pi\,\Psi_{{0}}-4\,\rho\,\Psi_{{1}}+3\,\kappa\,\Psi_{
{2}}+   \pic    \Phi_{{00}}+2\, 
\rhoc    \Phi_{{01}}+2\,\sigma\,\Phi_{{10}}\nonumber \\
-2\,\kappa\,\Phi_{{11}}-\kc    \Phi_{{02}}, \label{bi1}\\
\fl {\thd}   \Psi_{{0}}    -{\et}   \Psi_{{1}}
    +{\tho}   \Phi_{{02}}    -{\et}   \Phi_{
{01}}    =-\mu\,\Psi_{{0}}-4\,\tau\,\Psi_{{1}}+3\,\sigma\,\Psi_{
{2}}-\lc    \Phi_{{00}}+2\,   \pic    \Phi_{{01}}+2\,\sigma\,\Phi_{{11}}\nonumber \\
+   \rhoc    \Phi_{{02}}-2\,\kappa\,\Phi_{{12}}, \label{bi2}\\
\fl 3\,{\etd}   \Psi_{{1}}    -3\,{\tho}   \Psi_{{2}}
    +2\,{\tho}   \Phi_{{11}}    -2\,{\et}   
\Phi_{{10}}    +{\etd}   \Phi_{{01}}    -{\thd}
   \Phi_{{00}}    =3\,\lambda\,\Psi_{{0}}-9\,\rho\,\Psi_{{2
}}-6\,\pi\,\Psi_{{1}}+6\,\kappa\,\Psi_{{3}}\nonumber \\
+ (\muc    -2\,\mu )   \Phi_{{00}}+   2\,(\pi+
   \tauc  )      \Phi_{{01}}+2\,  ( \tau+   \pic  )      \Phi_{{10}}+2\,  ( 2\,
   \rhoc    -\rho )   \Phi_{{11}}\nonumber \\
   +2\,\sigma\,\Phi_{{20}
}-\sigmac    \Phi_{{02}}-2\,\kc    \Phi_{{12}}-2\,\kappa\,\Phi_{{21}}, \label{bi3}\\
\fl 3\,{\thd}   \Psi_{{1}}    -3\,{\et}   \Psi_{{2}}
    +2\,{\tho}   \Phi_{{12}}    -2\,{\et}   
\Phi_{{11}}    +{\etd}   \Phi_{{02}}    -{\thd}
   \Phi_{{01}}    =3\,\nu\,\Psi_{{0}}-6\,\mu\,\Psi_{{1}}-9
\,\tau\,\Psi_{{2}}+6\,\sigma\,\Psi_{{3}}\nonumber \\-\nuc
    \Phi_{{00}}+2\, (\muc  -\mu)    
    \Phi_{{01}}-2\,\lc    \Phi_{{10}
}+2\,  ( \tau+2 \pic)        \Phi_{{11
}}+   (2\,\pi+\tauc)        \Phi_{{02}}\nonumber \\
+   2\,(\rhoc    -\rho)    \Phi_{{12
}}+2\,\sigma\,\Phi_{{21}}-2\,\kappa\,\Phi_{{22}}. \label{bi4}
\end{eqnarray}

\section*{Acknowledgment}
All calculations were done using the Maple symbolic algebra system, while the properties of the metric (\ref{newmetric}) were checked with the aid of Maple's DifferentialGeometry
package\cite{Anderson_Torre}.\\

\end{document}